\documentclass[11pt]{article}
\usepackage{bm}% bold math
\usepackage{amsfonts}
\usepackage{amsgen}
\usepackage{amsbsy}
\usepackage{hyperref}% makes the references hyperlinked
\newcommand{\hodge}[1]{\,*#1}
\newcommand{\lie}{{\cal L}}
\begin{document}
\title{The horizon and its charges in the first order gravity}
\author{Rodrigo Aros\\ Universidad Nacional Andr\'{e}s Bello, Sazie 2320, Santiago,
Chile\\E-mail: raros@unab.cl}
\maketitle
\begin{abstract}
In this work the algebra of charges of diffeomorphisms at the horizon of generic black holes is
analyzed within first order gravity. This algebra reproduces the algebra of diffeomorphisms at the
horizon, (\emph{Diff} $(S^{1})$), without a central extension.
\end{abstract}

\section{Introduction}
In theory of fields the boundary conditions are essential not only to solve the differential
equations that arise from the action principle but also to define the intrinsic properties of a
theory. Actually, the action principle is not entirely defined by the bulk terms, it also needs
boundary terms (see for instance \cite{Regge:1974zd}). These ideas foresee that in a proper
quantum gravity theory boundary conditions should play an essential role.

In this line of arguments the formulation of the black holes entropy from first principles, as
elusive as it has been, for sure is to be related with the problem of boundary conditions (see for
instance \cite{Banados:1994qp,Brown:1993bq}). This have been actually proven in $2+1$ dimensions
where the entropy of the BTZ black hole \cite{Strominger:1998eq} can be computed in terms of the
central extension of the algebra of Hamiltonian charges of diffeomorphisms which preserve the
asymptotical boundary conditions \cite{Brown:1986nw}. Unfortunately the extension to higher
dimensions of those ideas can not be done verbatim, for instance, the algebra of the asymptotical
symmetries of the four dimensional asymptotical anti de Sitter spaces, which were studied in
\cite{Henneaux:1985tv}, is $SO(3,2)$, does not admit a non trivial central extension. Thus at
least a prescription for entropy in terms of a central extension concerning infinity does not
exist in any dimension.

In black holes physics the horizon can be regarded as an internal boundary, idea which for
instance allows to define the black holes thermodynamics in terms of N\"other charges
\cite{Wald:1993nt}. In this line in \cite{Carlip:1998wz,Carlip:1999cy} was argued that considering
the horizon as an internal boundary allows to attain an expression for the entropy of a black hole
in terms of the degeneracy of the diffeomorphisms preserving certain boundary conditions at the
horizon. An important issue, originally argued in \cite{Banados:1994qp}, corresponds to restrict
the discussion to the plane $(t,r)$ regardless the other directions. Obviously this result should
be valid in any dimension without further discussion. Even though some flaws have been reported in
original prescription \cite{Park:1999tj}, it seems that the idea, properly realized, actually
could represent a worth approach to the problem \cite{Dreyer:2001py,Park:2001zn}. In a similar
approach, but considering an induced conformal field theory near the horizon, in
\cite{Solodukhin:1998tc} was found out an expression for the entropy in terms of a central
extension of a Virasoro algebra.

The previous discussion is valid within the metric formalism of gravity, however gravity has an
alternative formulation, which is necessary when fermions are involved. This formulation is called
the first order formalism and is defined as follows: given a manifold $\mathcal{M}$ there exists
an orthonormal local basis of the (co)tangent space $e^{a}$ and a spin connection $\omega^{ab}$
which defines local Lorentz derivatives (see for instance \cite{gockeler}). In this formalism the
four dimensional Einstein Hilbert action with a negative cosmological constant reads
\begin{equation}\label{EH1}
  I_{EH} = \frac{1}{32\pi G} \int_{\mathcal{M}} R^{ab}\wedge e^c\wedge e^d \epsilon_{abcd}
   + \frac{1}{2l^2}e^a\wedge e^b \wedge e^c\wedge e^d \epsilon_{abcd},
\end{equation}
where $R^{ab}=d\omega^{ab} + \omega^a_{\,\,c}\omega^{cb}$ is the curvature two form which contains
the Riemann tensor as $R^{ab}=\frac{1}{2} R^{ab}_{\,\,\,cd} e^{c}\wedge e^{d}$,
$\Lambda=-3l^{-2}$, being $\Lambda$ the cosmological constant. From now on the differential forms
language will be assumed, thus $\wedge$ symbols are implicit.

The treatment of the horizon of a black hole in first order formalism can be cumbersome. The
essence of having a basis for the tangent space of a manifold specifically needs of a global
definition, otherwise topological considerations arise (See for instance
\cite{Epp:1998nm,Epp:1998cx}). Particularly if a global definition for the tangent space is
unattainable one should consider more than one fiber bundle and thus matching conditions between
them.

In this work only Euclidean manifolds are considered. In this case a black hole has no interior
region beyond the horizon, and it becomes the center of an Euclidean manifold. Also only suitable
for an Euclidean geometry vielbein's are considered, namely \textit{time} oriented ones
\footnote{That means that one of the orthonormal vectors is connected with the Wick rotated time
coordinate}. However in general any of these vielbeins become multivalued at the horizon, as it
happens in the center of any polar system of coordinates. Because of that in the first order
formalism one option it is to exclude the horizon,\emph{ i.e.}, to excise the point that it
represents by introducing a boundary around that point. This a way to circumvent the definition of
more than a single fiber bundle for the tangent space.

For a stationary spacetime the event horizon is a Killing horizon and thus it can be defined as
the hypersurface where the time like Killing vector, $\eta$, becomes light like. Rephrasing the
last paragraph, in terms of the vielbein the horizon is the hypersurface where $e^{a}(\eta)=0$ is
satisfied, yielding an undefined vierbein at the horizon. Because of that the horizon must be
removed out of the manifold, and thus it is introduced an internal boundary.

In this work the boundary conditions for the horizon an Euclidean black hole in four dimensions
within the first order formalism of gravity will be analyzed. Additionally, as it will be shown in
this work, it occurs that the temperature can be read from the relation $\omega^{01}(\eta) \propto
\beta^{-1}$ at the horizon \cite{Aros:2001gz} expressing that these boundary condition defines the
canonical ensemble. Essentially, it will be studied the diffeomorphisms which preserve a certain
set of boundary conditions at the horizon of a static black hole and the algebra they satisfy.
Particularly it will discussed a possible central extension of that algebra. The spacetime to be
discussed in this work is given by $\mathcal{M}=\mathbb{R}\times \Sigma$ where $\Sigma$
corresponds to a 3-dimensional spacelike hypersurface and $\mathbb{R}$ stands for the time
direction. The spacetime possesses an asymptotical locally AdS region, which defines the boundary
$\mathbb{R}\times\partial\Sigma_{\infty}$, and a horizon. As a matter of notation the boundary
involving the horizon will be denoted as $\mathbb{R}\times
\partial\Sigma_{H}$ thus one has that $\partial\mathcal{M}\mathbb{=R}\times
\partial\Sigma_{\infty} \cup \mathbb{R}\times \partial\Sigma_{H}$.

\section{First order formulation and boundary conditions}

Considering the problem of conserved charges in an asymptotically locally AdS (ALAdS) space, an
improved sound action principle for first order gravity was proposed in \cite{Aros:1999id}. The
action principle for (\ref{EH1}) is based on the fact that for any ALAdS space the Riemann tensor
behaves asymptotically as
\begin{equation}\label{Principle}
 R^{\mu\nu}_{\,\,\,\alpha\beta} \rightarrow -\frac{1}{l^{2}} \delta^{\mu\nu}_{\alpha\beta}.
\end{equation}
Therefore if Eq.(\ref{EH1}) is supplemented by the four dimensional Euler density\footnote{Euler
density is a closed form, thus its inclusion can not alter the field equations.},
\begin{equation}\label{EulerDensity}
  {\mathcal{E}}_{4} = \frac{l^{2}}{64\pi G}\int_{\mathcal{M}} R^{a b}\wedge R^{cd} \epsilon_{abcd},
\end{equation}
then variation of the new action $I=I_{EH}+{\mathcal{E}}_{4}$ is on shell a boundary term which
reads
\begin{equation}\label{BoundaryTerm}
    \delta I |_{\textrm{On shell}}=  \int_{\mathcal{\partial M}} \Theta(\delta\omega^{ab},e^{c}) =
    \frac{l^{2}}{32\pi G}\int_{\mathcal{\partial M}}\delta \omega^{ab}\bar{R}^{cd}\epsilon_{abcd},
\end{equation}
where $\bar{R}^{cd}=R^{cd}+ l^{-2} e^{c} e^{d}$. Using this result it is straightforward to prove
that there is no contribution from the asymptotic region $\mathbb{R}\times\partial\Sigma_{\infty}$
provided Eq.(\ref{Principle}). This last condition permits that the mass and angular momentum of
the Kerr-Neumann-AdS black hole to be computed as the N\"other charges associated with the time
and axial symmetries, respectively \cite{Aros:1999id}.

To consider the horizon as a boundary implies to set boundary conditions on it. For the action
$I=I_{EH}+{\mathcal{E}}_{4}$ to fix the spin connection on $\Sigma_{H}$ is an adequate boundary
condition. This was done in \cite{Aros:2001gz} demonstrating that the thermodynamics can be
obtained following this approach. In this work that condition will be relaxed.

\section{Local transformations}

A theory of gravity as (\ref{EH1}) is invariant under local Lorentz transformations and under
diffeomorphisms in the bulk, however the global analysis considering the boundary is subtler. The
transformations that preserve the boundary conditions, as occurs in $2+1$ dimensions, could give
rise on the border to dynamical degrees of freedom, even though in the bulk they represent
\emph{gauge} transformations merely.

Given that the vielbein is a basis, then any transformation of it can be written as a combination
of the basis itself, i.e., $\delta_{0}e^{a}=\Delta^{a}_{0\,\,b}e^{b}$ where $\Delta_{0}^{ab}$
depends on the transformation to be considered. Under a local Lorentz transformation the fields
change as
\[
\delta_{0} e^{a} = \lambda^{a}_{\,\,b}e^{b} \mbox{ and  }\delta_{0} \omega^{ab} =
-D(\lambda^{ab}),
\]
where $D$ is the Lorentz derivative and $\lambda^{ab}$ is a $0$-form antisymmetric Lorentz tensor,
\emph{i.e.}, it satisfies $\lambda^{ab}=-\lambda^{ba}$.

On the other hand, the transformation under diffeomorphisms is defined  by a vector $\xi$ as
$x'=x+\xi$. For any field $A$ it can be written in terms of a Lie derivative along $\xi$ as
$\delta_{0} A = -\lie_{\xi} A$. For the fields $(\omega^{ab},e^{a})$
\begin{eqnarray*}
&&\delta_{0}e^{a}=-\lie_{\xi} e^{a} = \Delta^{a}_{\,\,b}e^{b}\\
&&\delta_{0}\omega^{ab}=-\lie_{\xi} \omega^{ab} = -D(I_{\xi} \omega^{ab}) - I_{\xi} R^{ab}
\end{eqnarray*}
where $\Delta^{ab}= I_{\xi}\omega^{ab} - e^{a \mu} e^{b \nu} (\nabla_{\mu} \xi_{\nu})$. If $\xi$
is a Killing vector then $\Delta^{ab}$ is antisymmetric and $ \delta_{0}\omega^{ab}=-\lie_{\xi}
\omega^{ab} = -D(\Delta^{ab})$, therefore $\Delta^{ab}$ can be regarded as the parameter of a
local Lorentz transformation.

\section{Horizon boundary condition}
In order to boundary term Eq.(\ref{BoundaryTerm}) vanish, given that infinity has no contribution,
one requires that
\begin{equation}\label{Horizon}
 \delta_{0}\omega^{ab}\bar{R}^{cd}\epsilon_{abcd}|_{\mathbb{R}\times\partial\Sigma_{H}} \sim 0.
\end{equation}

This condition in principle restricts the variation of the spin connection. The two transformation
above, Lorentz and diffeomorphisms, differ in this case. Meanwhile for the Lorentz transformations
the boundary term Eq.(\ref{Horizon}) reads
\begin{equation}\label{Lorentz}
D(\lambda^{ab})\bar{R}^{cd}\epsilon_{abcd}= d(\lambda^{ab}\bar{R}^{cd}\epsilon_{abcd}),
\end{equation}
which vanishes upon integration, for diffeomorphisms the corresponding condition
\begin{equation}\label{Lie}
\lie_{\xi}\omega^{ab}\bar{R}^{cd}\epsilon_{abcd}|_{R\times\partial\Sigma_{H}} \sim 0,
\end{equation}
can not be trivially satisfied and restricts the form of the vector fields $\xi$. Because any
Killing vector defines a Lorentz transformation and so satisfies the boundary condition
(\ref{Horizon}) they must be excluded from the discussion.

\section{N\"other charges and central extension}

As mentioned above the variations under diffeomorphisms can be described in terms of Lie
derivatives, which together with the N\"other method imply that the current $\hodge{{\bf J}_\xi} =
\Theta + I_\xi{\bf L}$ satisfies $d(\hodge{{\bf J}_\xi})=0$. For the action
$I=I_{EH}+{\mathcal{E}}_{4}$, its current reads
\begin{equation}\label{NoetherCurrent}
 \hodge{{\bf J}_\xi} = -d ( I_{\xi}\omega^{ab}\bar{R}^{cd} \epsilon_{abcd}).
\end{equation}
From this expression a \textit{conserved charge} can be defined as the integral of
(\ref{NoetherCurrent}) on a spacelike surface $\Sigma$. However, only if $\xi$ is a global
symmetry of the solution, that is if $\xi$ is a Killing vector, the charges may represent mass or
angular momentum.

The formal connection of these N\"other charges and the Hamiltonian charges was analyzed in
\cite{Aros:2001gz}, proving that they indeed agree.

The variation of the Hamiltonian charges associated with diffeomorphisms can be obtained using the
covariant phase space method \cite{Lee:1990nz}, thus
\[ \delta Q_{\xi} = \int_{\partial\Sigma}
     -\delta (I_{\xi}\omega^{ab}\bar{R}^{cd} \epsilon_{abcd})  +
    I_{\xi}( \delta \omega^{ab}\bar{R}^{cd}\epsilon_{abcd}).
\]

At the horizon, $\partial\Sigma_{H}$, the second term vanishes provided Eq.(\ref{Horizon}) and
thus the charge can be integrated as
\[
 Q_{\xi} = -\int_{\partial\Sigma_{H}}
     I_{\xi}\omega^{ab}\bar{R}^{cd} \epsilon_{abcd}.
\]

Now that the Hamiltonian charge has been established the variation under another diffeomorphisms,
defined by $\eta$, which satisfies the boundary conditions (\ref{Lie}) can be computed.
Additionally to the condition Eq.(\ref{Lie}) it is necessary to impose that the Lie bracket of two
vectors, which satisfy the boundary conditions, satisfies the boundary conditions. Recalling that
$\delta_{\eta} = -\lie_{\eta}$ the variation reads
\begin{equation}\label{Variation}
 \delta_{\eta} Q_{\xi} = \int_{\partial\Sigma}
     \lie_{\eta} (I_{\xi}\omega^{ab}\bar{R}^{cd} \epsilon_{abcd})
    -I_{\xi}(\lie_{\eta}\omega^{ab}\bar{R}^{cd}\epsilon_{abcd}).
\end{equation}
Now, the variation can be expressed as $\delta_{\eta} Q_{\xi} = Q_{[\eta,\xi]}+  K(\eta,\xi)$ with
\begin{equation}\label{CentralCharge}
    K(\eta,\xi) = \int_{\partial\Sigma_{H}}   (I_{\xi}\omega^{ab}\lie_{\eta}\bar{R}^{cd} \epsilon_{abcd})
    +(\lie_{\eta}\omega^{ab}I_{\xi}\bar{R}^{cd}\epsilon_{abcd}),
\end{equation}
which represents an extension of the algebra of diffeomorphisms.

\section{Topological black holes}
A feature of a negative cosmological constant is that besides the asymptotical AdS spaces it also
allows the existence of asymptotically \textbf{locally} AdS spaces (ALAdS), and among them the
usually called topological black holes (see for instance \cite{Vanzo:1997gw,Brill:1997mf}).
Topological black holes exist in any dimensions higher than $3$ and as well as for many theories
of gravity. They are defined in terms of the vielbein
\begin{equation}\label{vielbein}
    e^{0}=f(r) dt,\textrm{  }e^{1}= \frac{1}{f(r)}dr,\textrm{  }e^{m} = r \tilde{e}^{m},
\end{equation}
and its associated torsion free connection
\begin{equation}\label{spinconnection}
\omega^{01}=\frac{1}{2}\frac{d}{dr}f(r)^{2} dt,\textrm{ }\omega^{1m}= f(r) \tilde{e}^{m},
  \textrm{ } \omega^{mn} =  \tilde{\omega}^{mn},
\end{equation}
where $\tilde{e}^{m}=\tilde{e}^{m}_{i}(y)dy^{i}$ and $\tilde{\omega}^{mn}$ are a vielbein and its
associated torsion free connection on the transverse section with $m=2\ldots d-1$. The $y^{i}$'s
are an adequate set of coordinates. For instance in four dimensions
$f(r)^{2}=\gamma+l^{-2}r^{2}-2m/r$ and $\tilde{R}^{mn}=\gamma e^{m}e^{n}$.

For the geometry described by Eqs(\ref{vielbein},\ref{spinconnection}) the Killing vector which
defines the event horizon can be written as $\eta=\partial_{t}$ and thus
$e^{a}(\eta)=f(r)\delta^{a}_{0}$. Consequently the asymptotical behavior near the horizon can be
defined in terms of the function $f(r)$. It must be stressed that because $e^{a}(\eta)$ has a
geometrical origin being a scalar under diffeomorphisms its definition is coordinate independent.
Actually since it vanishes at the horizon one can regard the horizon as a fixed point, elsewhere
however the expression of $e^{a}(\eta)$ can differ depending on the coordinate system, although
its geometrical interpretation remains the same. It is worth noting that the condition near the
horizon $e^{a}(\eta)\rightarrow0$ is satisfied by every \textit{time} oriented vierbein on an
Euclidean manifold, however since there is no other sensible vierbein on an Euclidean manifold
that is completely general. Let us recall, however, that the expression is to be understood in a
convergence process, namely that $e^{a}(\eta)\rightarrow 0$ but never really becomes null since
the horizon, $r=r_{+}$, is not on the manifold.

To pursue in the discussion it is necessary to perform a Wick rotation on time, \emph{i.e.},  to
replace the $t$ by $-i\tau$ where $\tau=[0,\beta^{-1}[$ and $\beta$ is the inverse of the
temperature of the black hole. Considering the ansatz for vector fields
\begin{equation}\label{SmoothVector}
    \xi = \xi^{\tau}(\tau,r)\partial_{\tau}+\xi^{r}(\tau,r)\partial_{r},
\end{equation}
the non trivially null Lie derivatives $\lie_{\xi}\omega^{ab}$ read
\begin{eqnarray*}
  \lie_{\xi} \omega^{01} &=& \frac{1}{2} \left[ \frac{d^{2}f(r)^{2}}{dr^{2}} \xi^{r} +
  \frac{df(r)^{2}}{dr}\frac{\partial\xi^{\tau}}{\partial\tau}
    \right]d\tau+ \frac{1}{2} \frac{df(r)^{2}}{dr}\frac{\partial\xi^{\tau}}{\partial r} dr\\
 \lie_{\xi} \omega^{1m} &=& \xi^{r} \frac{df(r)}{dr} \tilde{e}^{m}.
\end{eqnarray*}
Therefore the necessary components of $\bar{R}^{ab}$ are
\begin{eqnarray*}
  \bar{R}^{23} &=& \left(\frac{r^{2}}{l^{2}}+\gamma-f(r)^{2}\right) \tilde{e}^{2} \tilde{e}^{3}\\
   \bar{R}^{0n}&=& \left(\frac{r}{l^{2}}-\frac{1}{2}\frac{df(r)^{2}}{dr}\right) f(r) dt\, \tilde{e}^{n}.
\end{eqnarray*}

Imposing that the component of $\xi$ converge to a finite -asymptotical- value as it
\textit{approaches} the horizon \cite{note} allows to expand $\xi$ as
\begin{eqnarray*}
\xi^{\tau}(\tau,r) &\approx&  \xi_{0}^{\tau}(\tau) +  \xi_{1}^{\tau}(\tau)(r-r_{+})+ O((r-r_{+})^{2}),\\
\xi^{r}(\tau,r) &\approx&  \xi_{0}^{r}(\tau) +  \xi_{1}^{r}(\tau)(r-r_{+})+ O((r-r_{+})^{2}),
\end{eqnarray*}
and requiring additionally that the Lie bracket of two vector fields preserves the boundary
conditions (\ref{Lie}) yields that $\xi$ reads
\[ \xi\approx h(\tau)\partial_{\tau}+ \left(A + B(r-r_{+}
)\right)\frac{dh(\tau)}{d\tau}
\partial_{r} + O((r-r_{+})^{2}),
\]
where $h(\tau)$ is an arbitrary function of $\tau$ and $A,B$ are functions of $\beta$. Now, given
that $\tau$ is a periodic variable then $h(\tau)$ can be expanded in terms of a Fourier basis as
\[
h(\tau)= \sum_{n} h_{n} \exp(2i\pi\beta n \tau),
\]
and therefore $\xi$ can be expanded as $\xi \approx \sum_{n} h_{n} \xi_{n}$ with
\begin{equation}\label{xin}
\xi_{n}= \exp(2i\pi\beta n \tau)\left[\partial_{\tau}+ 2i\pi\beta n \left(A + B(r-r_{+} )\right)
\partial_{r}\right].
\end{equation}

Now, it is direct to show that the vectors $\hat{\xi}_{n}=\frac{1}{2i\pi\beta}\xi_{n}$ satisfy the
Virasoro algebra
\[
  [\hat{\xi}_{m},\hat{\xi}_{n}] = (m-n) \hat{\xi}_{m+n}.
\]

Starting with this result one can compute the extension of the algebra of diffeomorphisms
(\ref{CentralCharge}), in this case for two of the vector of the form (\ref{xin}), namely
$K(\hat{\xi}_{m},\hat{\xi}_{n})$. It can be readily shown that
 \[
K(\hat{\xi}_{m},\hat{\xi}_{n}) = 0\textrm{ for any value }m,n,
 \]
which implies that in this case no central extension exists.

\section{Discussion}

In this work was analyzed the consequences of a particular set of boundary conditions for the
first order action of gravity $I=I_{EH}+{\mathcal{E}}_{4}$. This analysis naturally leads to study
the algebra of charges of diffeomorphisms on the horizon. The boundary conditions (\ref{Lie}) are
enough to determine the form of the smooth vectors (\ref{xin}) which preserve them. These vectors
satisfy the algebra of Virasoro (\emph{Diff} $(S^{1})$) as expected, however the algebra of
charges of diffeomorphisms reproduce the algebras of the vector fields without any central
extension. This result does not yield an expression for the entropy in term of the central
extension as was done in \cite{Carlip:1998wz,Dreyer:2001py,Park:2001zn}. One possible obstruction
to a non trivial central extension may be to have considered only smooth vector fields. As argued
in \cite{Dreyer:2001py} if non smooth vectors were considered a non trivial central extension
could have attained and thus a prescription for the entropy. However that idea needs, in the case
for metric formalism, the use of \emph{stretched horizons}, whose development in this case for
first order gravity, is beyond the scope of this work. Nonetheless that seems to be an appealing
direction to continue with this investigation.

\section{Acknowledgments} I would like to thank Professor Jorge Zanelli for helpful comments. This
work is partially funded by grant DI 08-02 (UNAB).

\vspace{0.25 in}
%\bibliographystyle{unsrt}
%\bibliographystyle{jhep}
%\bibliography{myXbib}

\providecommand{\href}[2]{#2}\begingroup\raggedright\endgroup

\end{document}